\documentclass[cits]{PoS}

\usepackage{amsmath,amssymb}
\usepackage{amsxtra,mathrsfs}
\usepackage{multirow,multicol}
\usepackage{graphicx}

\title{Multiparton NLO corrections by numerical methods}

\ShortTitle{Multiparton NLO corrections by numerical methods}

\author{\speaker{S. Becker}, D. G\"otz, C. Reuschle, C. Schwan and S. Weinzierl\\
         Johannes Gutenberg University, Mainz\\
         \email{becker@thep.physik.uni-mainz.de\\
                goetz@uni-mainz.de\\
                reuschle@uni-mainz.de\\
                schwan@uni-mainz.de\\
                stefanw@thep.physik.uni-mainz.de}}

\abstract{
In this talk we discuss an algorithm for the numerical calculation of one-loop QCD amplitudes 
and present results at next-to-leading order 
for jet observables in electron-positron annihilation calculated with the above-mentioned method.
The algorithm consists of subtraction terms, approximating the soft, collinear and ultraviolet divergences of QCD one-loop amplitudes,
as well as a method to deform the integration contour for the loop integration into the complex plane to match Feynman's $i\delta$ rule. 
The algorithm is formulated at the amplitude level and does not rely on Feynman graphs. 
Therefore all ingredients of the algorithm can be calculated efficiently using recurrence relations.
The application of this method to the leading-colour contribution of $e^+ e^- \rightarrow n \text{ jets}$, with $n$ up to seven,
demonstrates the efficiency of the approach. 
}

\FullConference{ 10th International Symposium on Radiative Corrections (Applications of Quantum Field Theory to Phenomenology) - Radcor2011\\
September 26-30, 2011\\
Mamallapuram, India}

\begin{document}

\newcommand{\sla}{\!\!\!/}
\newcommand{\pslash}{\not{\hbox{\kern-2.3pt $p$}}}
\newcommand{\kslash}{\not{\hbox{\kern-2.3pt $k$}}}
\newcommand{\ga}{\gamma}
\newcommand{\de}{\delta}
\newcommand{\ro}{\rho}
\newcommand{\si}{\sigma}
\newcommand{\al}{\alpha}
\newcommand{\be}{\beta}
\newcommand{\la}{\lambda}
\newcommand{\ka}{\kappa}
\newcommand{\om}{\omega}
\newcommand{\eps}{\varepsilon}
\newcommand{\dquad}{\quad\quad}
\newcommand{\tquad}{\quad\quad\quad}
\newcommand{\HRule}{\rule{\linewidth}{0.5mm}}
\newcommand{\mgr}{m_{gr}}
\newcommand{\mf}{m_f}
\newcommand{\kbar}{\bar{k}}
\newcommand{\ME}{\mathbb{H}}
\newcommand{\Lag}{\mathcal{L}}
\newcommand{\pole}{\frac{1}{\epsilon}}
\newcommand{\scriptA}{\mathcal{A}}
\newcommand{\scriptG}{\mathcal{G}}
\newcommand{\scriptL}{\mathcal{L}}
\newcommand{\order}{\mathcal{O}}
\newcommand{\vektor}{\overrightarrow}
\newcommand{\real}{\mathbb{R}}
\newcommand{\gleich}{\,=\,}
\newcommand{\gleichnospace}{\!=\!}
\newcommand{\scriptl}{\ell}
\newcommand{\mtop}{m_t}
\newcommand{\muv}{\mu_{UV}}
\newcommand{\pone}{p_1}
\newcommand{\qhat}{\hat{q}}
\newcommand{\qbar}{\bar{q}}

\section{Introduction}

The development of a fully numerical algorithm to calculate multi-parton QCD amplitudes and collider observables 
at next-to-leading order (NLO) accuracy and the implementation of such an algorithm in a 
viable Monte Carlo matrix element generator is a rather involved subject. 
However, the need for generators of this kind is motivated by the need for accurate QCD background predictions to LHC physics. 
On inclusion of further Standard Model interactions such a generator can also be used to directly compute other relevant processes 
in addition to QCD background. 
Our focus lies upon the virtual part of the NLO calculation, i.e. on evaluating the one-loop integration numerically, 
where we employ and extend the ideas of the subtraction method to the virtual part 
\cite{Becker:2011vg,Becker:2010ng,Assadsolimani:2010ka,Assadsolimani:2009cz,Gong:2008ww,Anastasiou:2007qb,Nagy:2006xy,Soper:2001hu}. 
In this regard our algorithm is different from the commonly used approaches, 
based on cut techniques and generalised unitarity or on more traditional Feynman graph approaches 
\cite{Berger:2009zg,Ita:2011wn,Ellis:2009zw,Melia:2010bm,Bevilacqua:2010ve,Bevilacqua:2009zn,Bredenstein:2009aj,Frederix:2010ne,vanHameren:2010cp,Badger:2010nx,Cascioli:2011va}, 
but shows promising features especially for the implementation in a numerical program. 
The algorithm consists of local subtraction terms to subtract divergences arising 
from the soft, collinear and ultraviolet regions of the virtual part, which render the integrand finite in the respective regions, 
and of a method to deform the integration contour of the loop integration into the complex plane in order to circumvent 
the remaining on-shell singularities. 
It works on the level of colour-ordered primitive amplitudes, where we utilise recursive algorithms 
to compute the corresponding one-loop off-shell currents for the bare primitive amplitudes, and is therefore fast and easily implemented.

The local subtraction terms for the soft and the collinear regions are formulated directly on the amplitude level.
These subtraction terms are proportional to the corresponding Born amplitudes and are easily implemented. 
The local subtraction terms for the ultraviolet region are known to include only propagator and vertex corrections, 
where the corresponding graphs are expanded around a new ultraviolet propagator. 
The total local ultraviolet subtraction term is constructed recursively from the propagator and vertex subtraction terms.
Once the local  subtraction terms are applied to the bare integrand 
it can be integrated numerically in four dimensions in loop-momentum space. 
However, singularities still remain, since one or more of the propagators still may go on-shell for certain real values of the loop-momentum.
To avoid these singularities we deform the integration contour into the complex space. 
The contour deformation can be implemented in two ways, either
by a direct deformation of the loop four-momentum only 
or alternatively 
by introducing Feynman parameters and deforming the loop four-momentum as well as the corresponding Feynman parameters. 
The numerical loop integration is performed together with the integration over the phase-space of the external particles in one Monte Carlo integration.
The subtraction terms yield simple results upon analytic integration over the loop-momentum, 
and the resulting pole structures cancel exactly against the pole structures from the soft and collinear parts of the real emission contributions 
as well as of the ultraviolet countertem from renormalisation. 
Hence, the algorithm goes hand in hand with the usual subtraction method for the real emission contributions, 
where we employ Catani-Seymour dipole subtraction \cite{Catani:1997vz,Dittmaier:1999mb,Phaf:2001gc,Catani:2002hc,Weinzierl:2005dd}. 
As a proof of concept the algorithm has been tested so far for $e^+ e^- \rightarrow n \text{ jets}$, with $n$ up to seven, 
for the massless case in the large-$N_c$ limit.
Up to four jets we reproduce the known results for the respective jet rates with very good agreement. 
Increasing the number of jets up to seven shows a good scaling behaviour in CPU time with respect to the number of final state particles.

\section{Subtraction method}

The subtraction method is widely used to render the real emission part of a NLO calculation suitable for a numerical Monte Carlo integration. 
The contributions to an infrared-safe observable at next-to-leading order with $n$ final state particles can be written as
\begin{eqnarray}
 \langle O \rangle^{NLO} &=& \int\limits_{n+1}O_{n+1}d\sigma^{R}+\int\limits_{n}O_{n}d\sigma^{V}+\int\limits_{n}O_{n}d\sigma^{C}.
\end{eqnarray}
Here a rather condensed notation is used. $d\sigma^{R}$ denotes the real emission contribution, 
whose matrix elements are given by the square of the Born amplitudes with $(n+3)$ partons $|A_{n+3}^{(0)}|^{2}$. 
$ d\sigma^{V}$ denotes the virtual contribution, whose matrix elements are given by the interference term of the one-loop 
and Born amplitude $\Re(A_{n+2}^{(0)^{*}}A_{n+2}^{(1)})$ and $d\sigma^{C}$ denotes a collinear subtraction term, 
which subtracts the initial state collinear singularities. 
Each term is separately divergent and only their sum is finite. 
One adds and subtracts a suitably chosen piece to be able to perform the phase space integrations by Monte Carlo methods:
\begin{eqnarray}
 \langle O \rangle^{NLO} &=& \int\limits_{n+1}\left(O_{n+1}d\sigma^{R}-O_{n}d\sigma^{A}\right)+\int\limits_{n}\left(O_{n}d\sigma^{V}+O_{n}d\sigma^{C}+O_{n}\int\limits_{1} d\sigma^{A}\right).
\end{eqnarray} 
The first term $\left(O_{n+1}d\sigma^{R}-O_{n}d\sigma^{A}\right)$ is by construction integrable over the $\left(n+1\right)$-particle phase space and can be evaluated numerically. 
After integration of the subtraction term over the unresolved one-parton phase space the infrared divergences of the virtual contribution 
from the one-loop amplitude cancel with the infrared poles of the subtraction terms. 
Therefore the second term is also infrared finite and can be evaluated numerically, provided the analytical result of the one-loop amplitude is known.
\par
We extend this subtraction method to the virtual part such that we can evaluate the one-loop integral of the one-loop amplitude numerically. 
The renormalised one-loop amplitude is related to the bare amplitude by
\begin{eqnarray}
\mathcal{A}^{(1)}&=&\mathcal{A}_{\mathrm{bare}}^{(1)}+\mathcal{A}_{\mathrm{CT}}^{(1)},
\end{eqnarray}
where $\mathcal{A}_{\mathrm{CT}}^{(1)}$ denotes the ultraviolet counterterm from renormalisation.
The bare amplitude involves the loop integration
\begin{eqnarray}
\mathcal{A}_{\mathrm{bare}}^{(1)}&=&\int\frac{d^{D}k}{(2\pi)^{D}}\mathcal{G}_{\mathrm{bare}}^{(1)}.
\end{eqnarray}  
where $\mathcal{G}_{\mathrm{bare}}^{(1)}$ denotes the integrand of the bare one-loop amplitude.
We introduce subtraction terms which match locally the singular behaviour of the bare integrand:
\begin{eqnarray}
\mathcal{A}_{\mathrm{bare}}^{(1)}+\mathcal{A}_{\mathrm{CT}}^{(1)} &=& 
 \int\frac{d^{D}k}{(2\pi)^{D}}\left(\mathcal{G}_{\mathrm{bare}}^{(1)}-\mathcal{G}_{\mathrm{soft}}^{(1)}-\mathcal{G}_{\mathrm{coll}}^{(1)}
                                  -\mathcal{G}_{\mathrm{UV}}^{(1)}\right)
 \nonumber \\
 &&
 {}+\left(\mathcal{A}_{\mathrm{CT}}^{(1)}+\mathcal{A}_{\mathrm{soft}}^{(1)}+\mathcal{A}_{\mathrm{coll}}^{(1)}+\mathcal{A}_{\mathrm{UV}}^{(1)}\right)
\end{eqnarray}
Analogous to $\mathcal{G}_{\mathrm{bare}}^{(1)}$, the integrands of the subtraction terms $\mathcal{A}_{x}^{(1)}$ are denoted by $\mathcal{G}_{x}^{(1)}$, 
where $x$ is equal to $\mathrm{soft}$, $\mathrm{coll}$ or $\mathrm{UV}$. 
The expression in the first bracket is finite and can therefore be integrated numerically in four dimensions.
The integrated subtraction terms in the second bracket can be easily calculated analytically in $D$ dimensions.
The poles in the dimensional regularisation parameter of the integrated subtraction terms are cancelled by the corresponding poles from the ultraviolet counterterms, 
initial state collinear subtraction terms and the integrated real emission subtraction terms. 
\par
In analogy to the one-loop amplitude we can write $d\sigma^{V}=d\sigma_{\mathrm{CT}}+\int \frac{d^Dk}{(2\pi)^D}d\sigma_{\mathrm{bare}}^{V}$ and then the NLO contributions reads
\begin{eqnarray}
\langle O \rangle^{NLO}
 & = &
 \int\limits_{n+1}\left(O_{n+1}d\sigma^{R}-O_{n}d\sigma^{A}\right)
 +\int\limits_{n+\mathrm{loop}}\left(O_{n}d\sigma_{\mathrm{bare}}^{V}-O_{n}d\sigma^{A'}\right)
 \nonumber \\
&&{}+\int\limits_{n}\left(O_{n}d\sigma_{\mathrm{CT}}+O_{n}d\sigma^{C}+O_{n}\int d\sigma^{A}+O_{n}\int\limits_{\mathrm{loop}}d\sigma^{A'}\right).
\end{eqnarray}
In a condensed notation this reads
\begin{eqnarray}
\langle O \rangle^{NLO}&=& \langle O \rangle_{\mathrm{real}}^{NLO}+\langle O \rangle_{\mathrm{virtual}}^{NLO}+\langle O \rangle_{\mathrm{insertion}}^{NLO}.
\end{eqnarray}
Every single term is finite and can be evaluated numerically.

\section{Local infrared subtraction terms}

Amplitudes in QCD may be decomposed into group-theoretical factors (carrying the colour structures) multiplied by kinematic factors 
called partial amplitudes.
At the loop level partial amplitudes may further be decomposed into primitive amplitudes.
A few important properties of primitive amplitudes shall be given: 
Firstly, primitive amplitudes are gauge invariant. This is important for the formal proof of the method. 
Secondly, for a given number of external legs primitive amplitudes have a fixed cyclic ordering of the external legs 
and a definite routing of the external fermion lines through the loop. This ensures that each propagator in the loop is uniquely defined in type, 
be it a quark or a gluon/ghost propagator, and position.
Due to the fixed cycling ordering there are only $n$ different loop propagators occuring in a primitive amplitude with $n$ external legs.
\begin{figure}[ht]
\centering
\includegraphics[scale=0.8]{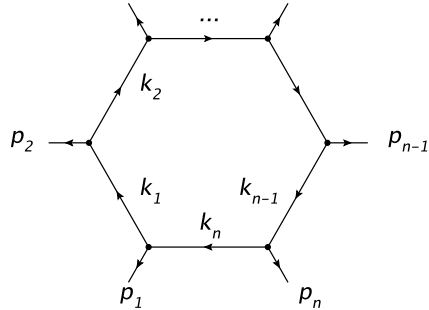}
\caption{
\label{fig_kinematics}
The labelling of the momenta for a primitive one-loop amplitude.}
\end{figure}
With the notation as in fig.~\ref{fig_kinematics} we define $k_{j} =  k-q_{j},$ with $ q_{j} =  \sum\limits_{i=1}^{j}p_{i}$,
where $k$ is the integration variable of the one-loop integral and the $p_{i}$'s are the external momenta. 
We can write the bare primitive one-loop amplitude as
\begin{eqnarray}
A_{\mathrm{bare}}^{(1)}&=&\int\frac{d^{D}k}{(2\pi)^{D}}G_{\mathrm{bare}}^{(1)},\qquad G_{\mathrm{bare}}^{(1)}\ = \ P_{a}(k)\prod\limits_{j=1}^{n}\frac{1}{k_{j}^{2}-m_{j}^{2}+i\delta}.
\end{eqnarray} 
$P_{a}(k)$ is a polynomial of degree $a$ in the loop momenta $k$ 
and the $+i\delta$-prescription tells us in which direction the poles of the propagators should be avoided.  

Soft singularities arise for $k \sim q_i$, whenever $p_i^2=m_{i-1}^2$, $m_i=0$, $p_{i+1}^2=m_{i+1}^2$. 
In this case we have a massless particle exchanged between two on-shell particles and the momentum $k_i$ is soft. 
Collinear singularities arise for $k \sim q_i-xp_i$, whenever $p_i^2=0$, $m_{i-1}=0$, $m_i=0$, where $x \in [0,1]$. 
In this case a massless external on-shell particle is attached to two massless propagators and the momenta $k_{i-1}$, 
$k_i$ and $p_i$ are collinear. 
The soft and collinear subtraction terms for massless QCD read
\begin{eqnarray}
G_{\mathrm{soft}}^{(1)} & = & 4i\sum\limits_{j\in I_g} \frac{p_{j}.p_{j+1}}{k_{j-1}^2k_{j}^2k_{j+1}^2} A_{j}^{(0)} \;, \nonumber \\
G_{\mathrm{coll}}^{(1)}
& = & -2i\sum\limits_{j\in I_g} \bigg[ \frac{S_{j}g_{\mathrm{UV}}(k_{j-1}^2,k_{j}^2)}{k_{j-1}^2k_{j}^2}
                                    + \frac{S_{j+1}g_{\mathrm{UV}}(k_{j}^2,k_{j+1}^2)}{k_{j}^2k_{j+1}^2} \bigg] A_{j}^{(0)} \;,
\end{eqnarray}
where the sum over $j \in I_g$ is over all gluon propagators $j$ inside the loop. 
Furthermore, $S_{j}=1$ if the external line $j$ corresponds to a quark 
and $S_{j}=1/2$ if it corresponds to a gluon. 
The function $g_{\mathrm{UV}}$ ensures that the integration over the loop momentum is ultraviolet finite.
The soft and collinear subtraction terms are formulated directly on the amplitude level and
are proportional to the corresponding Born amplitudes. 
Upon integration they yield simple analytic results:
\begin{eqnarray}
S_{\eps}^{-1}\mu_s^{2\eps}\int \frac{d^Dk}{(2\pi)^D} \, G_{\mathrm{soft}}^{(1)} & = &
-\frac{1}{(4\pi)^2}\frac{e^{\eps\gamma_E}}{\Gamma(1-\eps)}\sum\limits_{j\in I_g} \frac{2}{\eps^2} \Big( \frac{-2p_{j}p_{j+1}}{\mu_s^2}\Big)^{-\eps} A_j^{(0)} \; + {\cal O}(\eps),
 \nonumber \\
S_{\eps}^{-1}\mu_s^{2\eps}\int \frac{d^Dk}{(2\pi)^D} \, G_{\mathrm{coll}}^{(1)} & = &
-\frac{1}{(4\pi)^2}\frac{e^{\eps\gamma_E}}{\Gamma(1-\eps)}\sum\limits_{j\in I_g} (S_j+S_{j+1}) \frac{2}{\eps} \Big( \frac{\mu_{uv}^2}{\mu_s^2}\Big)^{-\eps} A_j^{(0)} \; + {\cal O}(\eps),
\end{eqnarray}
with $S_{\eps}\equiv(4\pi)^\eps\exp(-\eps\gamma_E)$ the typical volume factor in dimensional regularization, 
where $\gamma_E$ denotes the Euler-Mascheroni constant, $\mu$ denotes the renormalization scale in dimensional regularization 
and $\eps$ is defined through $D = 4-2\eps$.

The ultraviolet subtraction terms correspond to propagator and vertex corrections.
The subtraction terms are obtained by expanding the relevant loop propagators
around a new ultraviolet propagator $(\kbar^2-\mu_{uv}^2)^{-1}$, where $\kbar = k - Q$:
For a single propagator we have
\begin{eqnarray}
 \frac{1}{\left(k-p\right)^2}
 & = &
 \frac{1}{\bar{k}^2-\mu_{\mathrm{UV}}^2}
       + \frac{2\bar{k}\cdot\left(p-Q\right)}{\left(\bar{k}^2-\mu_{\mathrm{UV}}^2\right)^2}
 - \frac{\left(p-Q\right)^2+\mu_{\mathrm{UV}}^2}{\left(\bar{k}^2-\mu_{\mathrm{UV}}^2\right)^2}
 + \frac{\left[ 2\bar{k}\cdot\left(p-Q\right)\right]^2}{\left(\bar{k}^2-\mu_{\mathrm{UV}}^2\right)^3}
 + {\cal O}\left(\frac{1}{|\bar{k}|^5}\right).
\end{eqnarray}
We can always add finite terms to the subtraction terms.
For the ultraviolet subtraction terms we choose the finite terms such that
the finite parts of the 
integrated ultraviolet subtraction terms
are independent of $Q$ and proportional to the pole part, 
with the same constant of proportionality for all ultraviolet subtraction terms.
This ensures that the sum of all integrated UV subtraction terms is again proportional to a tree-level amplitude.

\section{Contour deformation}

Having a complete list of ultraviolet and infrared subtraction terms at hand, we can ensure that the integration
over the loop momentum gives a finite result and can therefore be performed in four dimensions.
However, this does not yet imply that we can safely integrate each of the four components of the loop momentum $k^\mu$
from minus infinity to plus infinity along the real axis.
There is still the possibility that some of the loop propagators go on-shell for real values of the loop momentum.
If the contour is not pinched this is harmless, as we may escape into the complex plane in a direction indicated by
Feynman's $+i\delta$-prescription.
However, it implies that the integration should be done over a region of real dimension $4$ in the complex space
${\mathbb C}^4$.
Let us consider an integral corresponding to a primitive one-loop amplitude with $n$ propagators minus the appropriate
IR- and UV-subtraction terms:
\begin{eqnarray}
\int\frac{d^{4}\tilde{k}}{(2\pi)^{4}}\left(\mathcal{G}_{\mathrm{bare}}^{(1)}-\mathcal{G}_{\mathrm{soft}}^{(1)}-\mathcal{G}_{\mathrm{coll}}^{(1)}-\mathcal{G}_{\mathrm{UV}}^{(1)}\right) &=& 
\int\frac{d^{4}\tilde{k}}{(2\pi)^{4}}P(\tilde{k})\prod\limits_{j=1}^{n}\frac{1}{\tilde{k}_{j}^{2}-m_{j}^{2}+i\delta}
\end{eqnarray}
where $P(\tilde{k})$ is a polynomial of the loop momentum $\tilde{k}^{\mu}$ and the integration is over a complex contour in order to avoid 
whenever possible the poles of the propagators. 
We discuss the method of the direct deformation of the loop momentum.
We set 
\begin{eqnarray}
\tilde{k}& = & k+i\kappa(k)
\end{eqnarray} 
where $k^{\mu}$ is real. After this deformation our integral equals
\begin{eqnarray}
&&\int\frac{d^{4}k}{(2\pi)^{4}}\left|\frac{\partial \tilde{k}^{\mu}}{\partial k^{\nu}}\right|P(\tilde{k}(k))\prod\limits_{j=1}^{n}\frac{1}{k_{j}^{2}-m_{j}^{2}-\kappa^{2}+2i k_{j}\cdot\kappa}.
\end{eqnarray}
To match Feynman's $+i\delta$-prescription
we have to construct the deformation vector $\kappa$ such that
\begin{eqnarray}
k_{j}^{2}-m_{j}^{2} & = & 0 \quad \rightarrow \quad k_{j}\cdot \kappa \geq 0. 
\end{eqnarray}
We remark that the numerical stability of the Monte Carlo integration depends strongly on the definition of the deformation vector $\kappa$.

\section{Recursion relations}

\begin{figure}[ht]
\begin{center}
\begin{minipage}{15.0mm}
\includegraphics[scale=0.9]{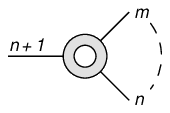}
\end{minipage}
\begin{minipage}{3.5mm}
{\normalsize{$=$}}
\end{minipage}
\begin{minipage}{5.0mm}
{\normalsize{$\sum\limits_{i=m}^{n-1}$}}
\end{minipage}
\begin{minipage}{14.0mm}
\includegraphics[scale=0.65]{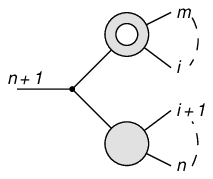}
\end{minipage}
\begin{minipage}{3.5mm}
{\normalsize{$+$}}
\end{minipage}
\begin{minipage}{5.0mm}
{\normalsize{$\sum\limits_{i=m}^{n-1}$}}
\end{minipage}
\begin{minipage}{14.0mm}
\includegraphics[scale=0.65]{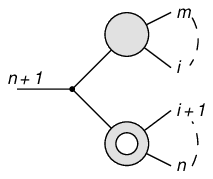}
\end{minipage}
\begin{minipage}{3.5mm}
{\normalsize{$+$}}
\end{minipage}
\begin{minipage}{20.0mm}
\includegraphics[scale=0.5]{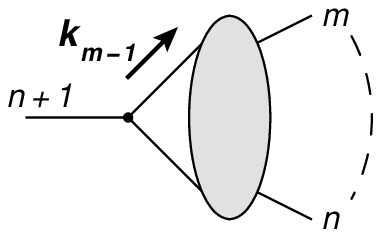}
\end{minipage}
\hspace*{-1mm} ,
\end{center}
\begin{center}
\hspace{0.75cm}
\begin{minipage}{15.0mm}
\includegraphics[scale=0.9]{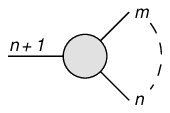}
\end{minipage}
\begin{minipage}{3.5mm}
{\normalsize{$=$}}
\end{minipage}
\begin{minipage}{5.0mm}
{\normalsize{$\sum\limits_{i=m}^{n-1}$}}
\end{minipage}
\begin{minipage}{15.0mm}
\includegraphics[scale=0.65]{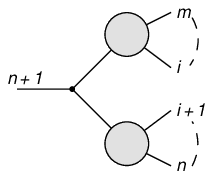}
\end{minipage}
\hspace*{-2mm} ,
\hspace{1.5cm}
\begin{minipage}{16.0mm}
\includegraphics[scale=0.8]{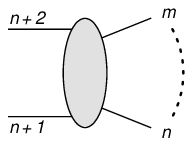}
\end{minipage}
\begin{minipage}{3.5mm}
{\normalsize{$=$}}
\end{minipage}
\begin{minipage}{10mm}
{\normalsize{$\sum\limits_{i=m-1}^{n-1}$}}
\end{minipage}
\begin{minipage}{18.0mm}
\includegraphics[scale=0.85]{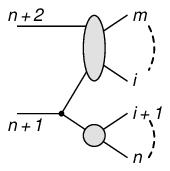}
\end{minipage}
\hspace*{-5mm} ,
\end{center}
\begin{center}
\begin{minipage}{15.0mm}
\includegraphics[scale=0.9]{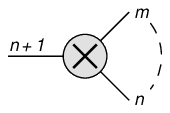}
\end{minipage}
\begin{minipage}{3.5mm}
{\normalsize{$=$}}
\end{minipage}
\begin{minipage}{10mm}
{\normalsize{$\sum\limits_{i=m}^{n-1}$}}$\;\;\Big($
\end{minipage}
\begin{minipage}{14.0mm}
\includegraphics[scale=0.65]{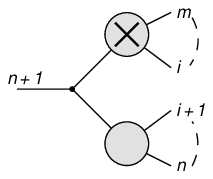}
\end{minipage}
\begin{minipage}{3.5mm}
{\normalsize{$+$}}
\end{minipage}
\begin{minipage}{14.0mm}
\includegraphics[scale=0.65]{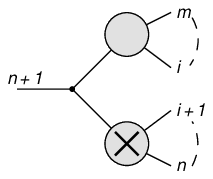}
\end{minipage}
\begin{minipage}{3.5mm}
{\normalsize{$+$}}
\end{minipage}
\begin{minipage}{15.0mm}
\includegraphics[scale=0.65]{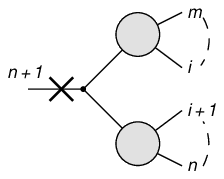}
\end{minipage}
\begin{minipage}{3.5mm}
{\normalsize{$+$}}
\end{minipage}
\begin{minipage}{15.0mm}
\includegraphics[scale=0.65]{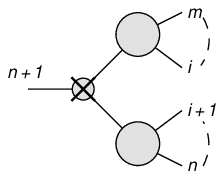}
\end{minipage}
\begin{minipage}{5mm}
$\Big)\;.$
\end{minipage}
\end{center}
\caption{\label{fig_recursion} Recursive relations for a three-valent toy model.}
\end{figure}

We use Berends-Giele type recursion relations \cite{Berends:1987me} 
to compute the tree amplitude, the bare one-loop integrand $G^{(1)}_{\mathrm{bare}}$ and the total UV subtraction term $G^{(1)}_{\mathrm{UV}}$.
These recursion relations are shown in fig.~\ref{fig_recursion} for the case of a three-valent toy model.

\section{NLO results for n-jets in electron-positron annihilation}

We have calculated results for jet observables in electron-positron annihilation, 
where the jets are defined by the Durham jet algorithm. 
The cross section for $n$ jets normalised to the LO cross section for $e^{+}e^{-}\rightarrow$ hadrons reads
\begin{eqnarray}
\frac{\sigma_{n-jet}(\mu)}{\sigma_{0}(\mu)}&=& \left(\frac{\alpha_{s}(\mu)}{2\pi}\right)^{n-2}A_{n}(\mu)+\left(\frac{\alpha_{s}(\mu)}{2\pi}\right)^{n-1}B_{n}(\mu)+\mathcal{O}(\alpha_{s}^{n}).
\end{eqnarray}
One can expand the perturbative coefficient $A_{n}$ and $B_{n}$ in $1/N_{c}$:
\begin{eqnarray}
A_{n}&=& N_{c}\left(\frac{N_{c}}{2}\right)^{n-2}\left[A_{n,\mathrm{lc}}+\mathcal{O}\left(\frac{1}{N_{c}}\right)\right],\qquad B_{n}\ = \ N_{c}\left(\frac{N_{c}}{2}\right)^{n-1}\left[B_{n,\mathrm{lc}}+\mathcal{O}\left(\frac{1}{N_{c}}\right)\right]
\end{eqnarray}
We calculate the leading order coefficient $A_{n,\mathrm{lc}}$ and the next-to-leading order coefficient $B_{n,\mathrm{lc}}$ for $n\leq 7$ 
at the renormalisation scale $\mu$ equal to the centre of mass energy. 
We take the centre of mass energy to be equal to the mass of the $Z$-boson. 
The scale variation can be restored from the renormalisation group equation. The calculation is done with five massless quark flavours.
\begin{figure}[ht]
\begin{center}
\includegraphics[bb= 125 460 490 710,width=0.32\textwidth]{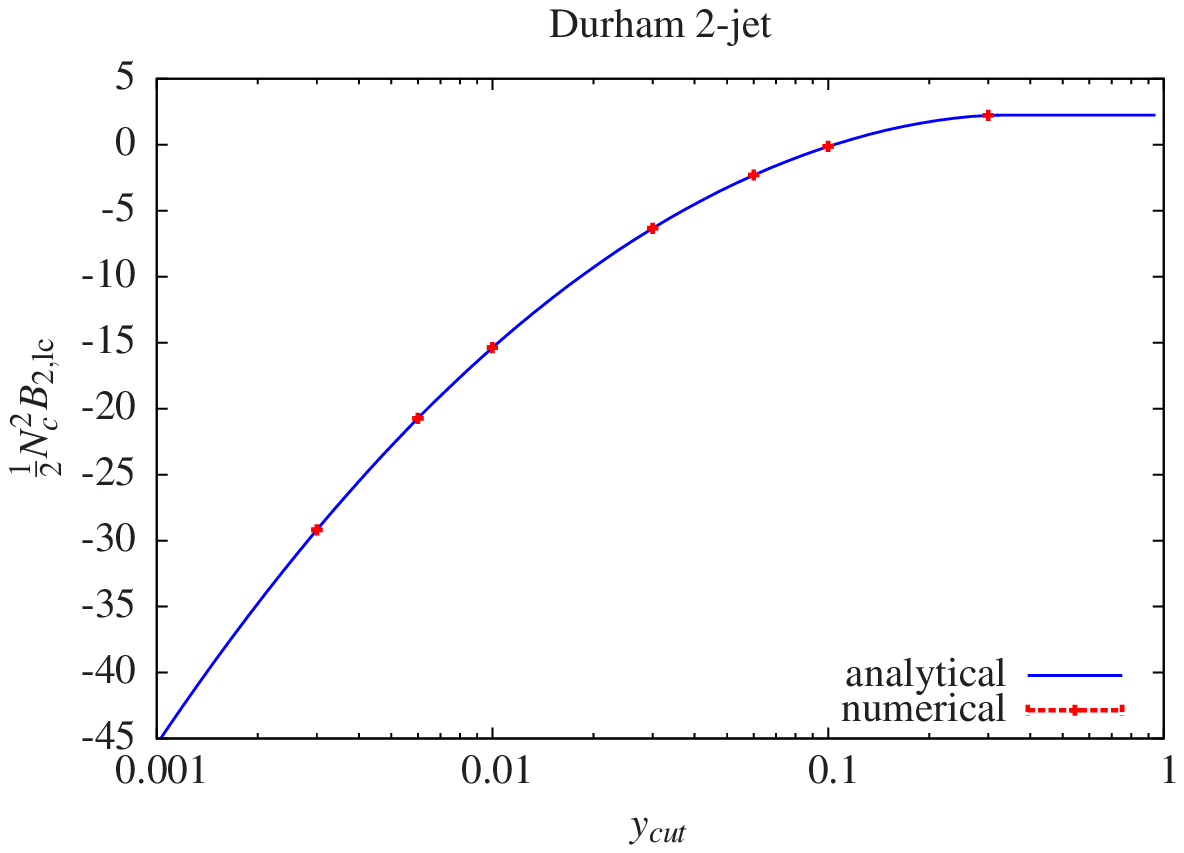}
\includegraphics[bb= 125 460 490 710,width=0.32\textwidth]{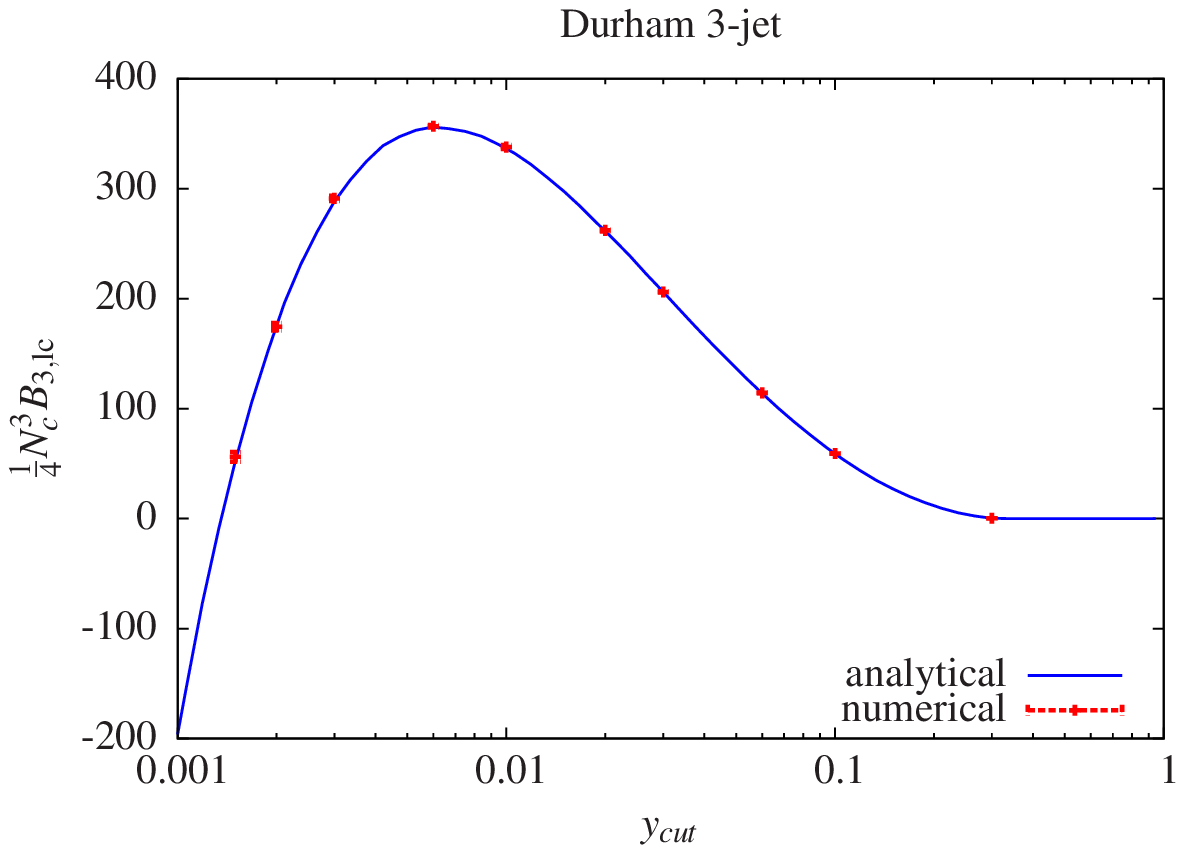}
\includegraphics[bb= 125 460 490 710,width=0.32\textwidth]{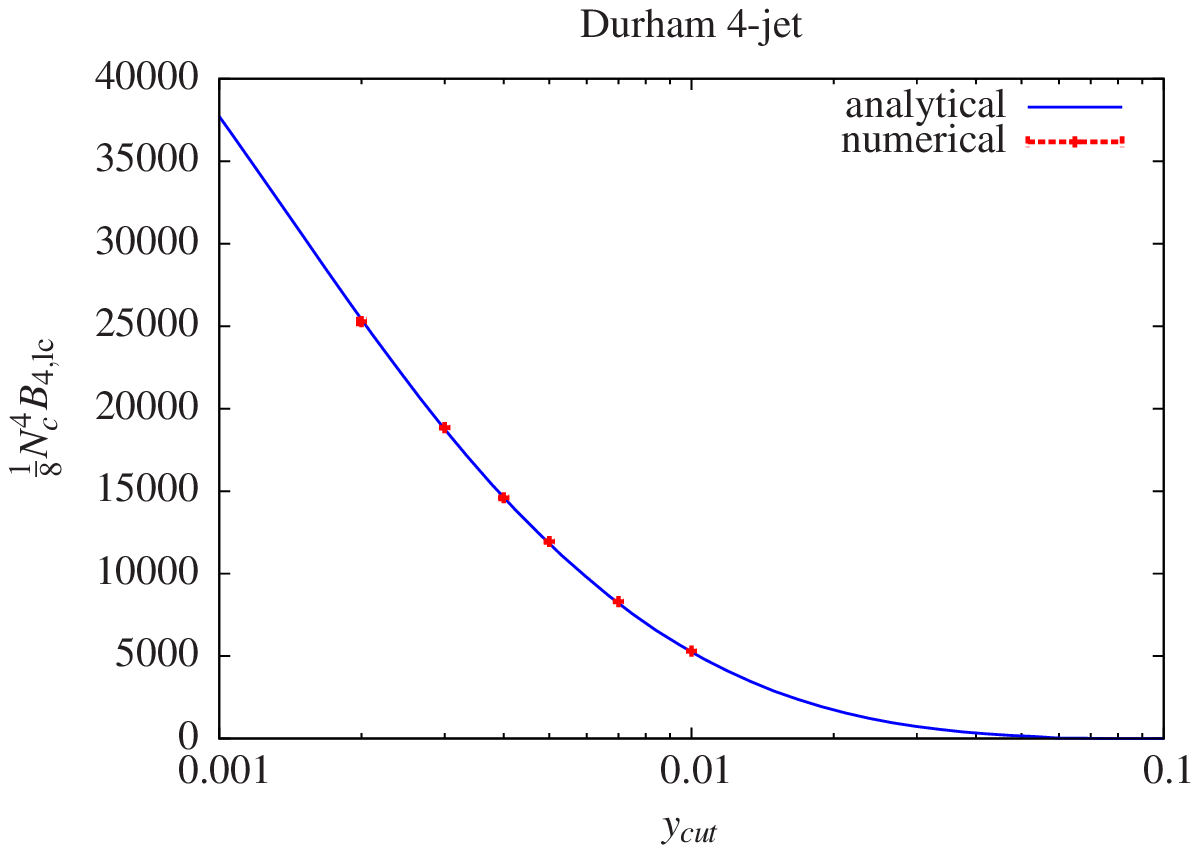}
\end{center}
\caption{
Comparison of the NLO corrections to the two-, three- and four-jet rate between the numerical calculation and an analytic calculation.
The error bars from the Monte Carlo integration are shown and are almost invisible.
}
\label{fig_jetrates}
\end{figure}
Fig.~\ref{fig_jetrates} shows the comparison of our numerical approach with the well-known results for two, three and four jets \cite{Weinzierl:1999yf,Weinzierl:2010cw}. 
We observe an excellent agreement. 
The results for five, six and seven jets for the jet parameter $y_{cut}=0.0006$ are
\begin{align}
 & 
 \frac{N_c^4}{8}  A_{5,\mathrm{lc}} = \left( 2.4764 \pm 0.0002 \right) \cdot 10^{4},
 & &
 \frac{N_c^5}{16} B_{5,\mathrm{lc}} = \left( 1.84 \pm 0.15 \right) \cdot 10^{6},
 \nonumber \\
 &
 \frac{N_c^5}{16} A_{6,\mathrm{lc}} = ( 2.874 \pm 0.002 ) \cdot 10^{5},
 & &
 \frac{N_c^6}{32} B_{6,\mathrm{lc}} = ( 3.88 \pm 0.18 ) \cdot 10^{7},
 \nonumber \\
 &
 \frac{N_c^6}{32} A_{7,\mathrm{lc}} = \left( 2.49 \pm 0.08 \right) \cdot 10^{6},
 & &
 \frac{N_c^7}{64} B_{7,\mathrm{lc}} = ( 5.4 \pm 0.3 ) \cdot 10^{8}.
\end{align}


\begin{thebibliography}{10}

\bibitem{Becker:2011vg}
S.~Becker, D.~Goetz, C.~Reuschle, C.~Schwan, and S.~Weinzierl,
\newblock (2011), arXiv:1111.1733.

\bibitem{Becker:2010ng}
S.~Becker, C.~Reuschle, and S.~Weinzierl,
\newblock JHEP {\bf 12}, 013 (2010), arXiv:1010.4187.

\bibitem{Assadsolimani:2010ka}
M.~Assadsolimani, S.~Becker, C.~Reuschle, and S.~Weinzierl,
\newblock Nucl. Phys. Proc. Suppl. {\bf 205-206}, 224 (2010), arXiv:1006.4609.

\bibitem{Assadsolimani:2009cz}
M.~Assadsolimani, S.~Becker, and S.~Weinzierl,
\newblock Phys. Rev. {\bf D81}, 094002 (2010), arXiv:0912.1680.

\bibitem{Gong:2008ww}
W.~Gong, Z.~Nagy, and D.~E. Soper,
\newblock Phys. Rev. {\bf D79}, 033005 (2009), arXiv:0812.3686.

\bibitem{Anastasiou:2007qb}
C.~Anastasiou, S.~Beerli, and A.~Daleo,
\newblock JHEP {\bf 05}, 071 (2007), hep-ph/0703282.

\bibitem{Nagy:2006xy}
Z.~Nagy and D.~E. Soper,
\newblock Phys. Rev. {\bf D74}, 093006 (2006), hep-ph/0610028;
%
\newblock JHEP {\bf 09}, 055 (2003), hep-ph/0308127.

\bibitem{Soper:2001hu}
D.~E. Soper,
\newblock Phys. Rev. {\bf D64}, 034018 (2001), hep-ph/0103262;
%
\newblock Phys. Rev. {\bf D62}, 014009 (2000), hep-ph/9910292;
%
\newblock Phys. Rev. Lett. {\bf 81}, 2638 (1998), hep-ph/9804454;

\bibitem{Berger:2009zg}
C.~F. Berger {\em et~al.},
\newblock Phys. Rev. Lett. {\bf 102}, 222001 (2009), arXiv:0902.2760;
%
\newblock Phys. Rev. {\bf D80}, 074036 (2009), arXiv:0907.1984;
%
\newblock Phys. Rev. {\bf D82}, 074002 (2010), arXiv:1004.1659;
%
\newblock Phys. Rev. Lett. {\bf 106}, 092001 (2011), arXiv:1009.2338.

\bibitem{Ita:2011wn}
H.~Ita {\em et~al.},
\newblock (2011), arXiv:1108.2229.

\bibitem{Ellis:2009zw}
R.~K. Ellis, K.~Melnikov, and G.~Zanderighi,
\newblock JHEP {\bf 04}, 077 (2009), arXiv:0901.4101;
%
\newblock Phys. Rev. {\bf D80}, 094002 (2009), arXiv:0906.1445.

\bibitem{Melia:2010bm}
T.~Melia, K.~Melnikov, R.~Rontsch, and G.~Zanderighi,
\newblock JHEP {\bf 12}, 053 (2010), arXiv:1007.5313.

\bibitem{Bevilacqua:2010ve}
G.~Bevilacqua, M.~Czakon, C.~G. Papadopoulos, and M.~Worek,
\newblock Phys. Rev. Lett. {\bf 104}, 162002 (2010), arXiv:1002.4009.

\bibitem{Bevilacqua:2009zn}
G.~Bevilacqua, M.~Czakon, C.~G. Papadopoulos, R.~Pittau, and M.~Worek,
\newblock JHEP {\bf 09}, 109 (2009), arXiv:0907.4723.

\bibitem{Bredenstein:2009aj}
A.~Bredenstein, A.~Denner, S.~Dittmaier, and S.~Pozzorini,
\newblock Phys. Rev. Lett. {\bf 103}, 012002 (2009), arXiv:0905.0110.

\bibitem{Frederix:2010ne}
R.~Frederix, S.~Frixione, K.~Melnikov, and G.~Zanderighi,
\newblock JHEP {\bf 11}, 050 (2010), arXiv:1008.5313.

\bibitem{vanHameren:2010cp}
A.~van Hameren,
\newblock Comput. Phys. Commun. {\bf 182}, 2427 (2011), arXiv:1007.4716.

\bibitem{Badger:2010nx}
S.~Badger, B.~Biedermann, and P.~Uwer,
\newblock Comput. Phys. Commun. {\bf 182}, 1674 (2011), arXiv:1011.2900.

\bibitem{Cascioli:2011va}
F.~Cascioli, P.~Maierhofer, and S.~Pozzorini,
\newblock (2011), arXiv:1111.5206.

\bibitem{Catani:1997vz}
S.~Catani and M.~H. Seymour,
\newblock Nucl. Phys. {\bf B485}, 291 (1997), hep-ph/9605323.

\bibitem{Dittmaier:1999mb}
S.~Dittmaier,
\newblock Nucl. Phys. {\bf B565}, 69 (2000), hep-ph/9904440.

\bibitem{Phaf:2001gc}
L.~Phaf and S.~Weinzierl,
\newblock JHEP {\bf 04}, 006 (2001), hep-ph/0102207.

\bibitem{Catani:2002hc}
S.~Catani, S.~Dittmaier, M.~H. Seymour, and Z.~Trocsanyi,
\newblock Nucl. Phys. {\bf B627}, 189 (2002), hep-ph/0201036.

\bibitem{Weinzierl:2005dd}
S.~Weinzierl,
\newblock Eur. Phys. J. {\bf C45}, 745 (2006), hep-ph/0510157.

\bibitem{Berends:1987me}
F.~A. Berends and W.~T. Giele,
\newblock Nucl. Phys. {\bf B306}, 759 (1988).

\bibitem{Weinzierl:1999yf}
S.~Weinzierl and D.~A. Kosower,
\newblock Phys. Rev. {\bf D60}, 054028 (1999), hep-ph/9901277.

\bibitem{Weinzierl:2010cw}
S.~Weinzierl,
\newblock Eur. Phys. J. {\bf C71}, 1565 (2011), arXiv:1011.6247.

\end{thebibliography}

\end{document}